\documentclass[conference ]{IEEEtran}
\usepackage{amsfonts}
\usepackage{times}
\usepackage{graphicx}
\usepackage{latexsym}
\usepackage{dsfont}
\usepackage{amssymb}
\usepackage{amsmath}
\usepackage{cite}
\usepackage{verbatim}
\usepackage{subfigure}

\def\bb0{{\mathbb{0}}}

\def\bb{{\mathbf{b}}}

\def\bg{{\mathbf{g}}}

\def\b0{{\mathbf{0}}}

\def\bX{{\mathbf{X}}}

\def\sf0{{\mathsf{0}}}

\usepackage{multirow}
\usepackage{multicol}
\usepackage{booktabs}
\usepackage{epstopdf}
\usepackage{enumerate}
\usepackage{algorithm}
\usepackage{amsmath}
\usepackage{float}
\usepackage{color}
\usepackage{makeidx}
\usepackage{bm}
\usepackage{cleveref}
\usepackage{url}
\usepackage{steinmetz}
\usepackage{varwidth}
\usepackage{soul}
\graphicspath{{figs/}}
\usepackage{tabularx}

\newcommand{\comm}[1]{}

\begin{document}

\title{Towards Real-World 6G Drone Communication: \\ Position and Camera Aided Beam Prediction }
\author{Gouranga Charan, Andrew Hredzak, Christian Stoddard, Benjamin Berrey, Madhav Seth, Hector Nunez, \\  Ahmed Alkhateeb\\ \textit{School of Electrical, Computer, and Energy Engineering, Arizona State University} \\ Emails: \{gcharan, ahredzak, cstodda2, bberrey,  mseth2, hanunez, alkhateeb\}@asu.edu }

\maketitle

\begin{abstract}

Millimeter-wave (mmWave) and terahertz (THz) communication systems typically deploy large antenna arrays to guarantee sufficient receive signal power. The beam training overhead associated with these arrays, however, make it hard for these systems to support highly-mobile applications such as drone communication. To overcome this challenge, this paper proposes a machine learning-based approach that leverages additional sensory data, such as visual and positional data, for fast and accurate mmWave/THz beam prediction. The developed framework is evaluated on a real-world multi-modal mmWave drone communication dataset comprising of co-existing camera, practical GPS, and mmWave beam training data. The proposed sensing-aided solution achieves a top-$1$ beam prediction accuracy of $86.32\%$ and close to $100\%$ top-3 and top-5 accuracies, while considerably reducing the beam training overhead. This highlights a promising solution for enabling highly mobile $6$G drone communications. 

\end{abstract}


\section{Introduction} \label{sec:Intro}

Future wireless communication systems in 5G-advanced, 6G, and beyond will need to reliably support highly-mobile devices such as drones and autonomous vehicles. Drones (and unmanned aerial vehicles (UAVs)) \cite{bariah2020drone} are envisioned to form the basic building block of next-generation aerial networks and are the key to enabling futuristic applications such as extending the coverage of mmWave/sub-THz wireless networks, supporting latency-critical applications, and enhancing the capabilities of security monitoring systems. To satisfy the high data rate requirements of these novel applications, the drones will need to be equipped with mmWave/THz \cite{Rappaport2019} transceivers and deploy large antenna arrays. Carefully adjusting the narrow beams of these arrays at both the transmitters and receivers is essential to guarantee sufficient receive SNR. Adjusting these narrow beams, however, is typically associated with large training overhead which scales with the number of antennas. Furthermore, the three-dimensional motion along with the highly mobile nature of the drone necessitates a frequent update to the optimal beam index, which further increases the beam training overhead. This motivates looking for new approaches to overcome the challenges and enable highly-mobile mmWave/THz drone communication.

In recent years, several solutions \cite{Hur2011,Alkhateeb2014,Jayaprakasam2017,Alkhateeb2018,Khan2020a,Alrabeiah_camera} have been developed to overcome the beamforming/channel estimation overhead. Initial approaches focused on three main directions: (i) Beam training with adaptive beam codebook \cite{Hur2011,Alkhateeb2014}, (ii) compressive channel estimation by leveraging channel sparsity \cite{Alkhateeb2014}, and (iii) designing beam tracking solutions \cite{Jayaprakasam2017}. In beam training, the optimal beam at the transmitter and receiver is obtained using exhaustive or adaptive beam training, incurring a large beam-training overhead and is not suitable for highly-mobile multi-user scenarios \cite{Hur2011,Alkhateeb2014}. By leveraging the inherent sparsity of mmWave channels, \cite{Alkhateeb2014} formulates the mmWave channel estimation as a sparse reconstruction problem. Although the compressive channel estimation techniques help in reducing the beam training overhead, they can typically save only one order of magnitude in the training overhead. Further, for these solutions, the training overhead scales with the number of antennas, reducing the impact for systems with large antenna arrays. Next, \cite{Jayaprakasam2017} proposed an extended Kalman filter-based (EKF) channel tracking solution to maintain the communication link between the basestation and mobile user. Although such an EKF-based beam tracking approach helps in minimizing the beam training overhead, they can only predict beams for a short future time window and their performance is normally limited in NLOS scenarios. The limitations of these approaches motivate the development of more efficient beam prediction approaches.

Leveraging machine learning (ML) to address the beam prediction task has gained increasing interest in the last few years \cite{Alkhateeb2018, RobertPos,Elizabeth, Charan22,GCharanBlockage}. These solution mainly focus on leveraging the additional information to provide awareness about the wireless environment. In \cite{Alkhateeb2018}, the authors propose to utilize the receive wireless signature to predict the optimal beam indices at the basestation. Such a solution, though promising, is limited to a single-user setting. Position information was leveraged in \cite{RobertPos,Elizabeth} to predict the optimal beam index. Although the solutions can help in reducing the training overhead, relying only on location alone might result in inaccurate predictions due to the inherent errors associated with the GPS data. In \cite{Charan22,GCharanBlockage}, we proposed to leverage the visual data (captured by cameras)to predict the optimal beam indices. These solutions, however, are based on synthetic data and focused on scenarios with humans, vehicles, or robots as the transmitter, where the users typically move in easy to predict mobility patterns in two dimensions. In general, vehicles or robots tend to travel in the azimuthal plane without any change in the elevation during movement. Drones or UAVs have six degrees of freedom, three translation, and three rotation, which further increases the challenge of predicting the optimal beam index. An important question that arises is whether the promising results in \cite{Charan22,GCharanBlockage} can be achieved in reality for mmWave drones?

In this paper, we attempt to answer this important question. In particular, we propose a deep learning-based sensing-aided solution to reduce the beam training overhead in mmWave/THz drone communication. The main contribution of this work can be summarized as follows:
\begin{itemize}
	\item {Formulating the sensing-aided beam prediction problem for mmWave/THz drone communication considering practical visual and communication models.}
	\item{Developing a novel deep learning-based solution for mmWave/THz drone beam prediction by utilizing different sensory data such as vision (captured at the basestation) and the position, orientation, and height of the drone. }
	\item{Providing the first real-world evaluation of sensing-aided drone beam prediction based on our large-scale dataset, DeepSense 6G \cite{DeepSense}, that consists of co-existing multi-modal sensing and wireless communication data.}
\end{itemize}
Based on the adopted real-world dataset, the developed solution achieves $\approx 86\%$ top-1 (and close to 100$\%$ top-3) beam prediction accuracy. This highlights the capability of the proposed sensing-aided beam prediction approaches in significantly reducing the beam training overhead.

\begin{figure}[!t]
	\centering
	\includegraphics[width=0.8\linewidth]{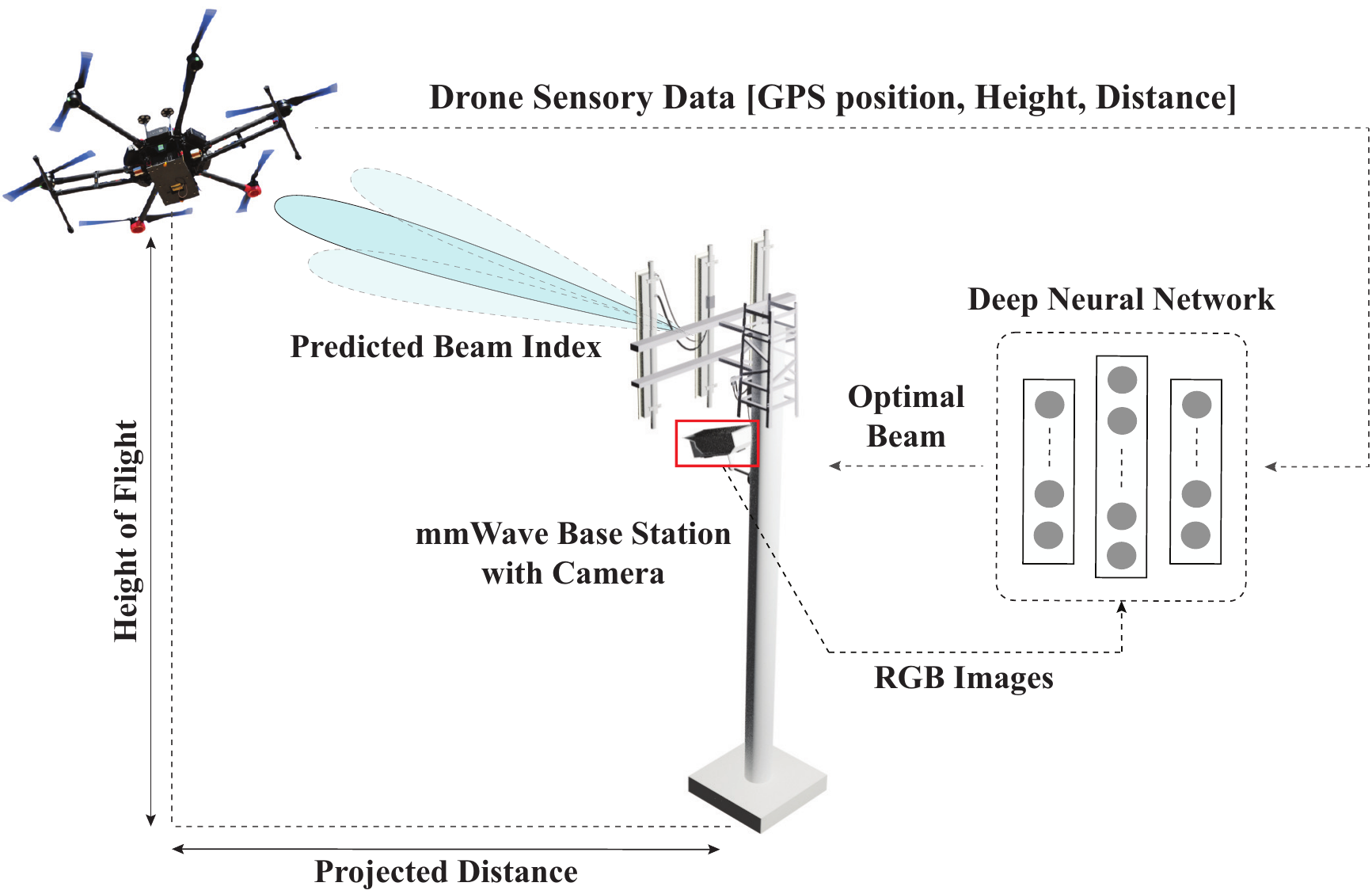}
	\caption{An illustration of the mmWave basestation serving a drone in a real wireless environment. The basestation utilizes additional sensing data such as RGB images, GPS location of the drone, etc., to predict the optimal beam.}
	\label{fig:beam_pred_key_idea}
	\vspace{-2mm}
\end{figure}

\section{Sensing-Aided Beam Prediction: \\ System Model and Problem Formulation }\label{sec:sys_ch_mod}
This work considers a communication system where a mmWave basestation is serving a drone flying at different speeds and heights in a real wireless communication environment. In this section, we first present the adopted wireless communication system model. Then, we formulate the sensing-aided beam prediction problem.

\subsection{System Model} \label{sec:sys_model}

This paper adopts the system model illustrated in Fig.~\ref{fig:beam_pred_key_idea}, where a basestation, equipped with an $M$-element uniform linear array (ULA) and an RGB camera, is serving a flying drone. The drone carries a single-antenna transmitter and is equipped with a GPS receiver capable of collecting real-time position information. The adopted communication system employs OFDM transmission with K subcarriers and a cyclic prefix of length D. To serve the mobile user, the basestation is assumed to employ a pre-defined beamforming codebook $\boldsymbol{\mathcal F}=\{\mathbf f_q\}_{q=1}^{Q}$, where $\mathbf{f}_q \in \mathbb C^{M\times 1}$ and $Q$ is the total number of beamforming vectors. In the downlink transmission (from the basestation to the drone), if $\mathbf h_{k}[t] \in \mathbb C^{M\times 1}$ denotes the channel between the basestation and the drone at the $k$th subcarrier and time $t$, then the received signal at the drone can be written as 
\begin{equation}\label{eq:sys_mod}
	y_{k}[t] = \mathbf h_{k}^T[t] \mathbf f_q[t]x + v_k[t],
\end{equation}
where $\mathbf f \in \boldsymbol{\mathcal F}$ is the optimal beamforming vector at time $t$ and $v_k[t]$ is a noise sample drawn from a complex Gaussian distribution $\mathcal N_\mathbb C(0,\sigma^2)$. The transmitted complex symbol $x\in \mathbb C$ need to satisfy the following constraint $\mathbb E\left[ |x|^2 \right] = P$, where $P$ is the average symbol power. The beamforming vector $\mathbf f^{\star}[t] \in \boldsymbol{\mathcal F}$ at each time step $t$ is selected to maximize the average receive SNR and is defined as 
\begin{equation}\label{eq:beam_training}
	\mathbf f^{\star}[t] = \underset{\mathbf f_q[t]\in \mathcal F}{\text{argmax}} \frac{1}{K}\sum_{k=1}^{K} \mathsf{SNR}|\mathbf h_{k}^T[t] \mathbf f_q[t] |^2,
\end{equation}
where $\mathsf{SNR}$ is the transmit signal-to-noise ratio, SNR = $\frac{P}{\sigma^2}$. 


\begin{figure*}[!t]
	\centering
	\includegraphics[width=0.9\linewidth]{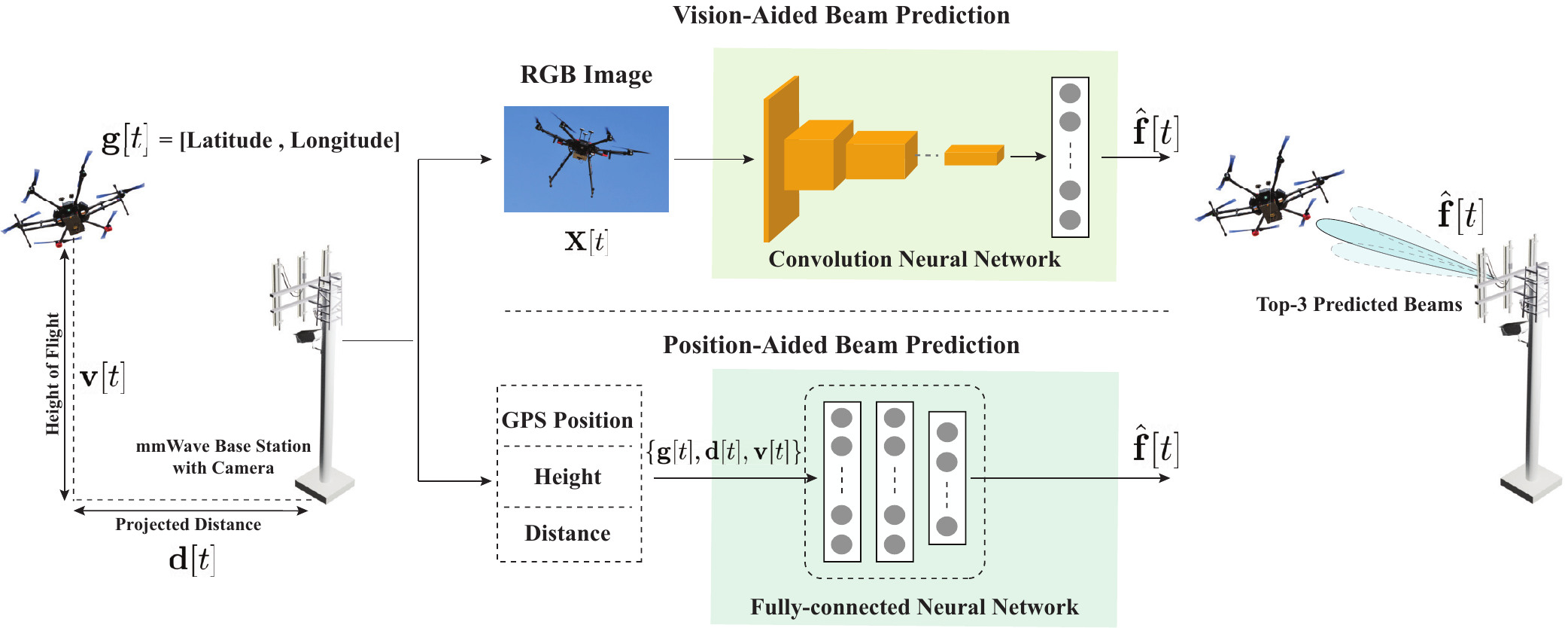}
	\caption{A block diagram showing the proposed solution for both the vision and position-aided beam prediction task. As shown in the figure, the camera installed at the basestation captures real-time images of the drone in the wireless environment. A CNN is then utilized to predict the optimal beam index. The basestation receives the information for the other three sensing data, which is then provided to a fully-connected neural network to predict the beam.  }
	\label{fig:beam_pred_soln}
\end{figure*}

\subsection{Problem Formulation} \label{sec:prob_form}

Given the system model in Section~\ref{sec:sys_model}, if the basestation wants to select an optimal beam $\mathbf f^{\star}[t]$ out of its codebook $\boldsymbol{\mathcal F}$ to serve the drone, then it can determine this optimal beam that maximizes the received power based on \eqref{eq:beam_training}. The optimum beam is computed by either utilizing the explicit channel knowledge which is hard to acquire in mmWave/THz systems or by performing an exhaustive search over the beam codebook, which is typically associated with high beam training overhead. This makes it challenging for mmWave/THz systems to support the highly-mobile drones. In this paper, instead of following the conventional beam training approach, \textbf{we propose to predict the optimal beam index for the transmitter by utilizing the sensory data (position or vision) collected by the basestation or the drone.} In particular, this work assumes the availability of the following sensory data at the basestation: (i) the RGB images captured by a camera installed at the basestation, (ii) the GPS positional data collected by the drone and fed back to the basestation, and (iii) the height/distance of the drone in the real wireless environment. Formally, we define $\bg[t] \in \mathbb R^2$  as the two-dimensional position vector of the transmitter (consisting of the latitude and longitude information) at time step $t$. And we define $\bX[t] \in \mathbb{R}^{W \times H \times C}$ as the corresponding RGB image, captured by a camera installed in the basestation at time $t$, where $W$, $H$, and $C$ are the width, height, and the number of color channels of the image. Further, let $d[t] \in \mathbb R^1$ and $v[t] \in \mathbb R^1$ denote the height and the distance of the transmitter from the stationary unit at time instance $t$. The objective of the drone beam prediction task is to find a prediction/mapping function $f_{\Theta}$ that utilizes the available sensory data, $\mathcal{S}[t] = \left\lbrace \bg[t], \bX[t], d[t], v[t]\right\rbrace$ to predict (estimate) the optimal beam index $ \hat{\mathbf f}[t] \in \boldsymbol{\mathcal F}$ with high fidelity. The mapping function can be formally expressed as
\begin{equation}
	f_{\Theta}: \mathcal{S}[t] \rightarrow  \hat{\mathbf f}[t].
\end{equation}
In this work, we develop a machine learning model to learn this prediction function $f_{\Theta}$. Let $ \mathcal D = \left\lbrace \left (\mathcal{S}_u, \mathbf f^{\ast}_u \right) \right\rbrace_{u=1}^U $ represent the available dataset consisting of sensing data-beam pairs is collected from the real wireless environment, where $U$ is the total number of samples in the dataset. Then, the objective is to maximize the number of correct predictions over all the sample in $ \mathcal D$. This can be formally written as 
\begin{equation}\label{eq:prob_form_1}
	f^{\star}_{\Theta^{\star}} = \underset{f_{\Theta}}{\text{argmax}}\\ \prod_{u=1}^U \mathbb P\left( \hat{\mathbf f}_u = \mathbf f^{\star}_u | \mathcal{S}_u \right),
\end{equation}
where the product in \eqref{eq:prob_form_1} is due to the implicit assumption that the samples in the dataset $ \mathcal D$ are drawn from an independent and identically distribution (i.i.d.). The prediction function is parameterized by a set $\Theta$ representing the model parameters and learned from the dataset $ \mathcal D$ of labeled data samples. Next, we present our proposed machine learning model for sensing-aided mmWave/THz drone beam prediction.

\section{Sensing-Aided Beam Prediction: \\ A Deep Learning Solution} \label{sec:prop_sol}
In this section, we present an in-depth overview of the proposed beam prediction solution. First, we present the key idea in Section~\ref{sec:key_idea} and then explain the details of our proposed solution in Section~\ref{sec:ml_model}.

\subsection{Sensing-Aided Drone Beam Prediction: Key Idea}\label{sec:key_idea}

The mmWave/THz communication systems require large antenna arrays and use narrow directive beams to guarantee sufficient signal power gain. This is primarily to overcome the severe path loss associated with the high-frequency signals. Selecting the optimal beams in these systems is typically associated with large beam training overhead, which becomes more challenging in high-mobility dynamic wireless environments with moving transmitters, reflectors, and scatters. The highly mobile nature with very high flying speeds of the drones and the added capability of hovering or traveling in a three-dimensional space further increases the challenges faced in the mmWave drone communication system. Instead of relying on conventional beam training, this work selects the optimal beam index by utilizing additional sensory data. In this paper, the task of selecting the optimal beam index from a pre-defined codebook, $\boldsymbol{\mathcal F}$,  at any coherence time, is defined as the \textbf{beam prediction} task.

\begin{figure*}[t]
	\centering
	\subfigure[]{\centering \includegraphics[width=0.31\linewidth]{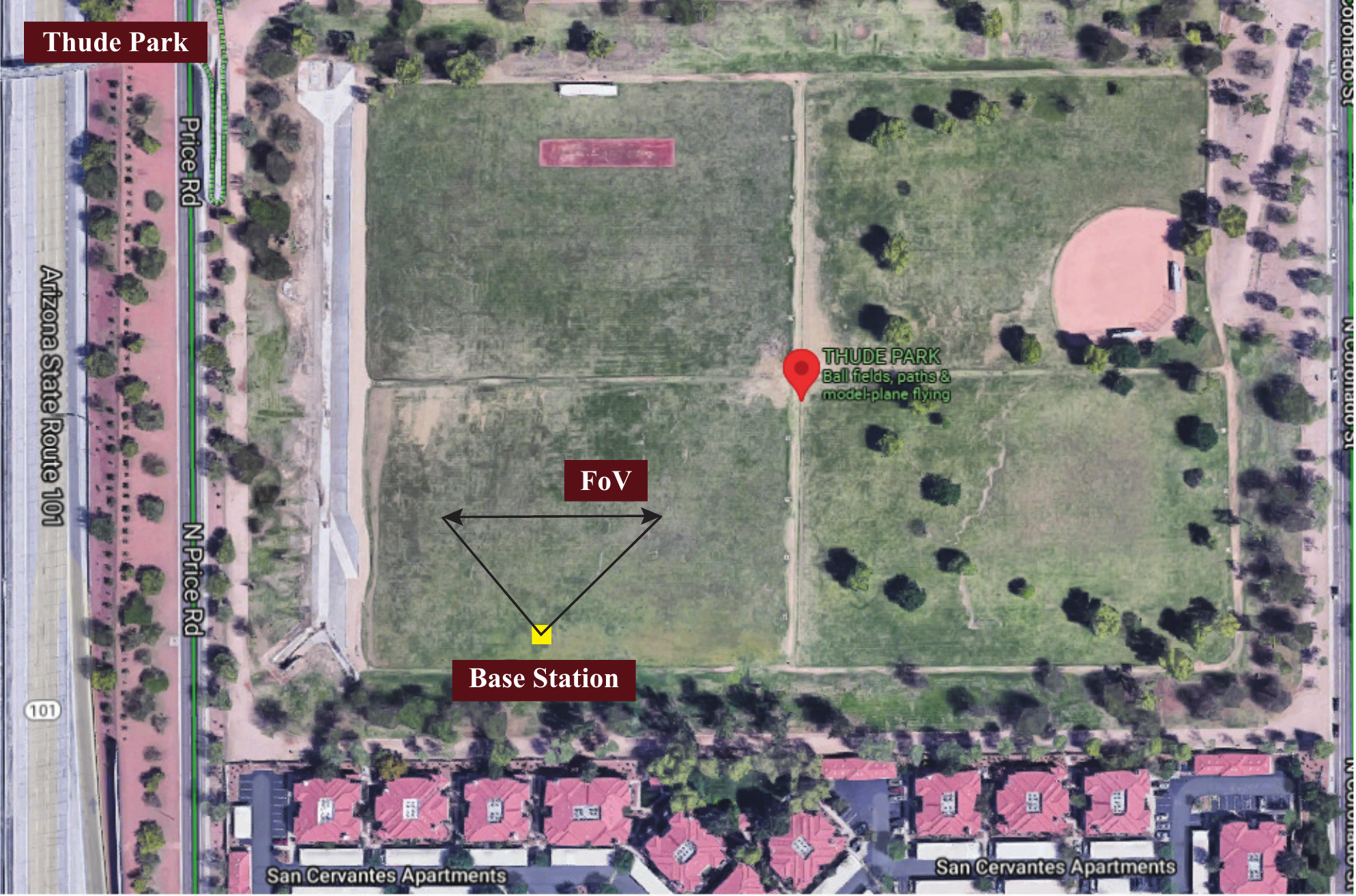}\label{fig:thude_park_map}}
	\subfigure[]{\centering 
	\includegraphics[width=0.31\linewidth]{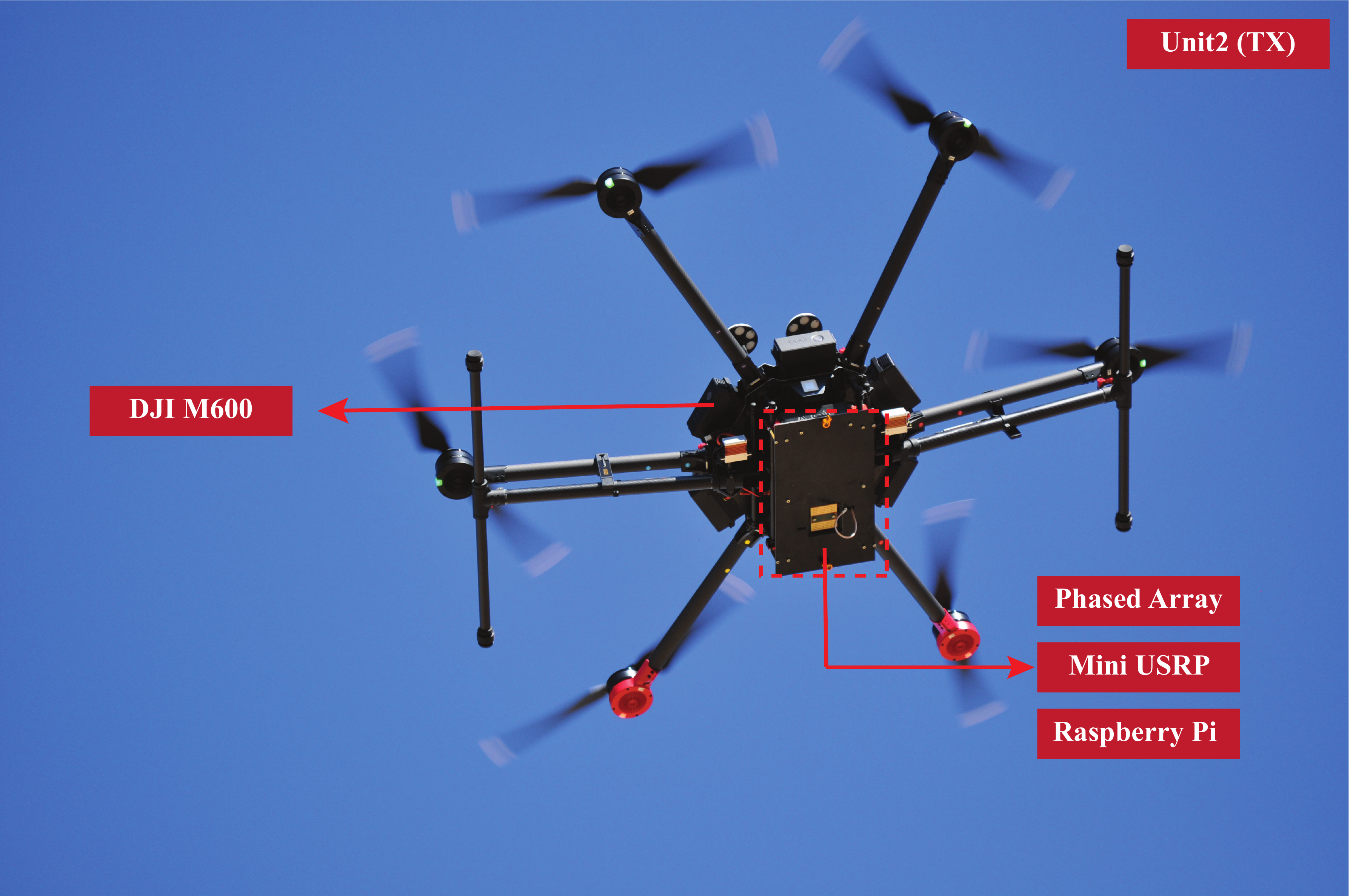}\label{fig:drone_view1}}
	\subfigure[]{\centering \includegraphics[width=0.31\linewidth]{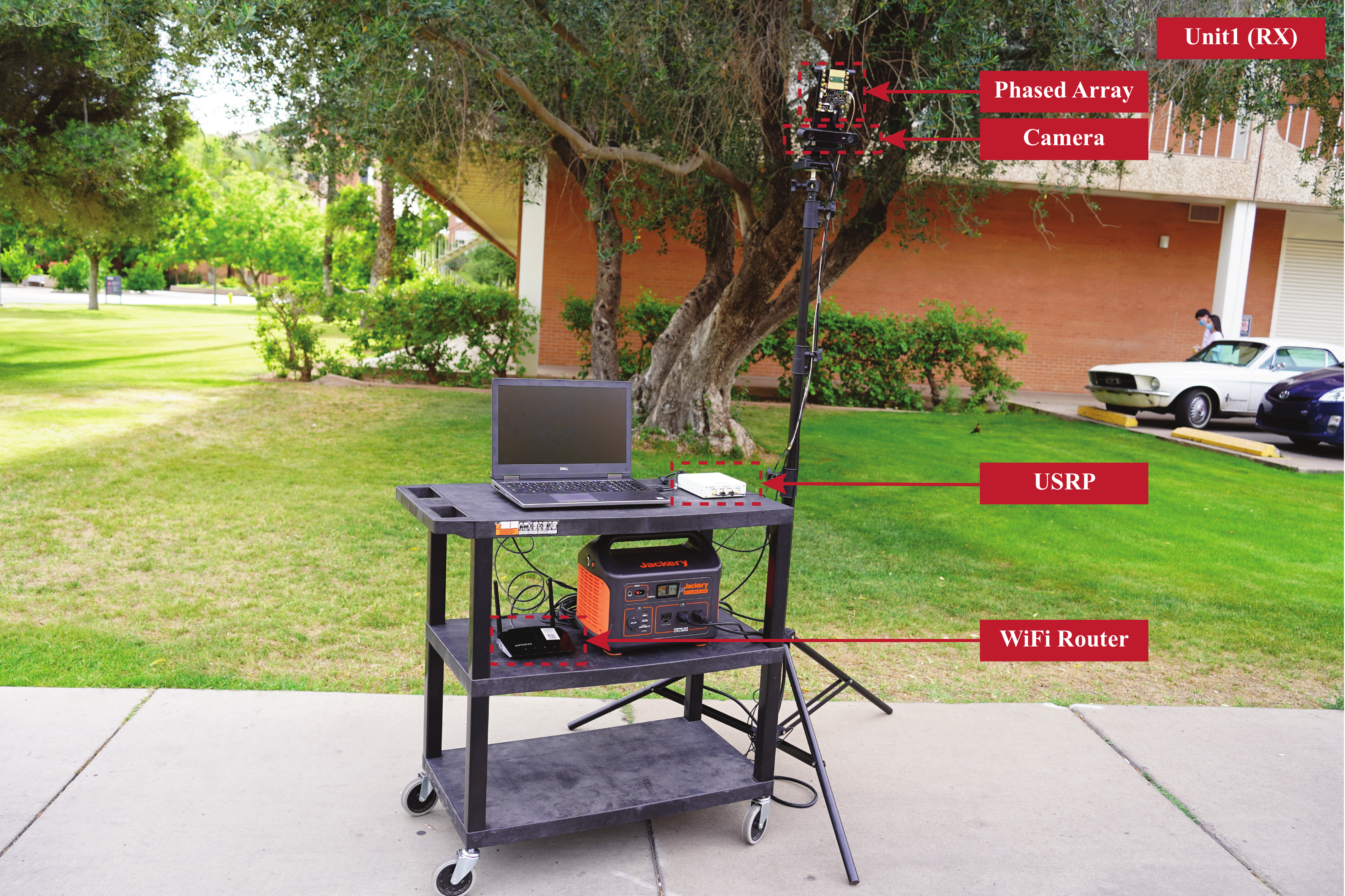}\label{fig:basestaion}}
	\caption{This figure presents the overview of the DeepSense 6G testbed and the location used in this scenario. (a) Shows the google map top-view of Thude Park utilized for this data collection. Fig. (b) and (c) present the different components of the drone (acting as the transmitter) and the basestation. In (b), we highlight the mmWave phased array attached to the drone transmitting signals to the basestation on the 60 GHz band.}
	\label{fig:deepsense}
\end{figure*}

High-frequency systems suffer from large path-loss, which makes line-of-sight (LOS) a preferable setting. This dependence of both high-frequency communication on LOS operation forms the building block of the sensing-aided solution. In general, the beamforming vectors provide directional information that summarizes the dominant signal direction for well-calibrated antenna arrays. The beam vectors divide the scene (spatial dimensions) into multiple (possible overlapping) sectors, where each sector is associated with a particular beam value. Therefore, given a pre-defined codebook, the beam prediction task can be transformed into a classification task, where depending on the user location in the wireless environment, a beam index from the codebook is assigned. With this motivation, we plan to utilize visual and positional data to predict the optimal beam indices. The recent advancements in computer vision and object detection have provided the capability to accurately detect the different objects and extract the user's relative position in the visual scene. Similarly, for any outdoor location, advanced positioning systems such as GPS can be used to accurately (with some error margin) locate a user in the scene. At every time step, the basestation captures a visual image of its environment and receives the sensing data such as GPS location, the height, and distance of the transmitter. Instead of performing beam training at every step, the basestation utilizes machine learning models and the additional sensory data to predict the optimal beamforming vector from a pre-defined codebook.

\subsection{Proposed Solution}\label{sec:ml_model}
In this section, we describe the proposed machine learning-based solutions for sensing-aided beam prediction. First, we present a detailed overview of the proposed position-aided beam prediction task followed by the details of the vision-aided beam prediction task. In Fig.~\ref{fig:beam_pred_soln}, we present the block diagram of the proposed beam prediction ML model. 

\subsubsection{Proposed Position-aided Solution} \label{sec:position_beam_pred}

This sub-task entails the prediction of optimal beam index by leveraging the positional information of the transmitter. A Multi-Layer Perceptron (MLP) network is adopted to perform the position-aided beam prediction task. The inputs to the MLP network are the normalized Latitude and Longitude values. The MLP network is designed to have two hidden layers with $512$ hidden units each and an output layer consisting of $Q$ units. ReLU activation function is applied to the output of the hidden layers in order to introduce non-linearity. Since the position-aided beam prediction task is posed as a classification problem as presented in Section~\ref{sec:prob_form}, the softmax function is applied to the final output layer. 

In order to perform a comparative evaluation of the different sensing modalities, we extend the solution to utilize the other sensing data, such as the height and the distance of the drone from the basestation. For this, along with the normalized GPS data, we provide the normalized height and distance information to the proposed ML model. The rest of the architecture is similar to the solution proposed for the position-alone beam prediction.

\subsubsection{Proposed Vision-aided Solution} \label{sec:vision_beam_pred}

In this subsection, we present our proposed deep learning model for the vision-aided beam prediction task. The objective is to learn the prediction function $f_{\Theta}(\bX[t])$ by utilizing only visual data. The ideal choice of the deep learning model for this task, as mentioned above, is the CNNs. The idea is to utilize CNN to perform this classification task, i.e., the model learns to map an image to a beam index. The CNN in the proposed solution needs to meet two essential requirements: (i) an accurate and generalizable classifier for the vision-based classification task and (ii) a low computational footprint. The residual neural network (ResNet) \cite{resnet} has proven to be highly efficient for image classification tasks and, most importantly, addresses the two major requirements mentioned above. For this particular task, a ResNet-50 \cite{resnet} model has been selected as the primary choice of CNN. However, instead of training the ResNet model from scratch on the single candidate beam prediction dataset, we select an ImageNet2012 pre-trained ResNet model as the initial architecture. The model is further modified by removing the last fully connected layer and replacing it with a layer consisting of $Q$ neurons. The fully connected layer parameters are initialized randomly following a normal distribution with zero mean and unit variance. Unlike conventional transfer learning, the ResNet-50 model is fine-tuned end-to-end in a supervised fashion, using a labeled dataset.


\section{Testbed Description and Development Dataset}\label{sec:datset}

In order to evaluate the performance of the proposed sensing-aided mmWave drone prediction solution, we utilize the \textbf{DeepSense 6G} \cite{DeepSense} dataset. DeepSense 6G is a real-world multi-modal dataset dedicated to sensing-aided wireless communication applications. It contains co-existing multi-modal data such as vision, mmWave wireless communication, GPS data, LiDAR, and Radar collected in a real-wireless environment. This section presents a brief overview of the scenario adopted from the DeepSense 6G dataset, followed by the analysis of the final development dataset utilized for the sensing-aided beam prediction study.



\subsection{DeepSense 6G: [Scenario 23]}  \label{sec:testbed}

This study adopts Scenario 23 of the DeepSense 6G dataset specifically designed to study high-frequency wireless communication applications with drones. The hardware testbed and the exact location used for collecting these data are shown in Fig.~\ref{fig:deepsense}. The DeepSense testbed $4$ is utilized for this data collection consisting of a stationary and a mobile unit. The testbed is deployed at the Southwest corner of the park, as shown in Fig.~\ref{fig:thude_park_map}. The stationary unit \{unit1 (RX)\} is equipped with a standard-resolution RGB camera, and mmWave Phased array. The stationary unit adopts a 16-element ($M = 16$) 60GHz-band phased array and it receives the transmitted signal using an over-sampled codebook of $64$ pre-defined beams ($Q = 64$). The mmWave phased array and the RGB camera is placed on a table at the height of $\approx 1.5$ meters from the ground level. Both the camera and the phased array are facing towards the sky, which helps increase the basestation's field-of-view (FoV). In this data collection scenario, the mobile unit \{unit2 (RX)\} is the RC drone equipped with a mmWave transmitter, GPS receiver, and inertial measurement units (IMU).  The transmitter consists of a quasi-omni antenna constantly transmitting (omnidirectional) at 60 GHz band. In order to increase the diversity of the dataset, the drone is flown at varying heights and distances from the basestation with different speeds of flight. For more information regarding the data collected testbed and setup, please refer to \cite{DeepSense}.

\begin{table}[!t]
	\caption{DeepSense 6G Scenario 23: Development Dataset}
	\centering
	\setlength{\tabcolsep}{5pt}
	\renewcommand{\arraystretch}{1.3}
	\begin{tabular}{c|c|cc}
		\hline \hline
		\multirow{2}{*}{\textbf{Task}}                                                                     & \multirow{2}{*}{\textbf{Modality}}                                                   & \multicolumn{2}{c}{\textbf{Number of Samples}}              \\ \cline{3-4} 
		&                                                                                      & \multicolumn{1}{c|}{\textbf{Training}} & \textbf{Test} \\ \hline \hline
		\multirow{4}{*}{\textbf{\begin{tabular}[c]{@{}c@{}}Sensing-Aided \\ Beam Prediction\end{tabular}}} & \textbf{GPS Position}                                                                & \multicolumn{1}{c|}{8402}              & 3602                \\ \cline{2-4} 
		& \textbf{GPS Position + Height}                                                       & \multicolumn{1}{c|}{8402}              & 3602                \\ \cline{2-4} 
		& \textbf{\begin{tabular}[c]{@{}c@{}}GPS Position \\ + Height + Distance\end{tabular}} & \multicolumn{1}{c|}{8402}              & 3602                \\ \cline{2-4} 
		& \textbf{RGB Image}                                                                   & \multicolumn{1}{c|}{8402}              & 3602                \\ \hline \hline
	\end{tabular}
	\label{ref:tab_dataset_analysis}
\end{table}



\begin{table}[!t]
	\caption{Beam Prediction: Design and Training Hyper-parameters}
	\centering
	\setlength{\tabcolsep}{5pt}
	\renewcommand{\arraystretch}{1.2}
	\begin{tabular}{@{}l|cc@{}}
		\toprule
		\toprule
		\textbf{Parameters}                     & \textbf{Vision}  & \textbf{Position/Combined }            \\ \midrule \midrule
		\textbf{ML Model}                       & ResNet-50           & 2-layered MLP                   \\
		\textbf{Batch Size}                     & 32                  & 32                                 \\
		\textbf{Learning Rate}                  & $1 \times 10 ^{-4}$ & $1 \times 10 ^{-2}$ \\
		\textbf{Learning Rate Decay}            & epochs 4, 8 and 12       & epochs 20, 40 and 80  \\
		\textbf{LR Reduction Factor} & 0.1                 & 0.1                               \\
		\textbf{Total Training Epochs}          & 20                  & 100                              \\ \bottomrule \bottomrule
	\end{tabular}
	\label{tab_beam_pred_train_params}
	\vspace{-4mm}
\end{table}


\subsection{DeepSense 6G: [Development Dataset]} \label{sec:dev_data} 

The evaluation of the proposed sensing-aided beam prediction solution requires data collected in a real wireless environment with a drone as the mmWave transmitter. In this work, we utilize the publicly available scenario 23 of the DeepSense 6G dataset. The different data modalities collected are the RGB images, real-time GPS location, distance, height, speed, and orientation of the drone, and a $64 \times 1$ vector of mmWave received power. The first step involves downsampling the $64 \times 1$ power vector to $32 \times 1$ ($Q = 32$) by selecting every alternate sample in the vector. Since the basestation receives the mmWave signal using an oversampled codebook of $64$ pre-defined beams, the downsampling does not affect the total area covered by the beams. In order to compute the updated optimal beam index for a particular sample, we select the index of the beam with maximum received power in the downsampled power vector. The final step in the processing pipeline is dividing the dataset into training and test sets following a 70-30\% split. In Table~\ref{ref:tab_dataset_analysis}, we present the details of the development datasets for the sensing-aided beam prediction task.

\section{Performance Evaluation} \label{sec:perf_eval}
This section studies the performance of the proposed solutions for the sensing-aided beam prediction task. In the first sub-section, we will present the details of the experimental setup. Next, we discuss the performance of the proposed solution for the different sub-tasks presented in Section~\ref{sec:prop_sol}.

\subsection{Experimental Setup:} \label{sec:exp_setup}

\textbf{Network Training:} In this work, we propose to utilize different sensing data modalities to perform the beam prediction task, i.e., position-alone, position and height combined, position, height and distance combined, and visual data. As presented in Section~\ref{sec:prop_sol}, different modality-specific deep learning models are proposed to perform the sensing-aided beam prediction task. For the position-alone and the combined data modalities, we develop 2-layered fully-connected neural networks. For the vision-aided approach, the proposed solution adopts a ResNet-50 model to predict the optimal beam indices. The proposed ML models are trained and validated on the task-specific dataset as presented in Section~\ref{sec:dev_data}. The cross-entropy loss with the Adam optimizer is used to train the models. The details of the hyper-parameters used to fine-tune the models are presented in Table~\ref{tab_beam_pred_train_params}.

\textbf{Evaluation Metric:} The primary metric adopted to evaluate the proposed solution is the top-k accuracy. Note that the top-k accuracy is defined as the percentage of the test samples where the optimal ground-truth beam is within the top-$k$ predicted beams. This work presents the top-1, top-2, top-3, and top-5 accuracies to evaluate the proposed solutions comprehensively. 

\begin{figure}[!t]
	\centering
	\includegraphics[width=1.0\linewidth]{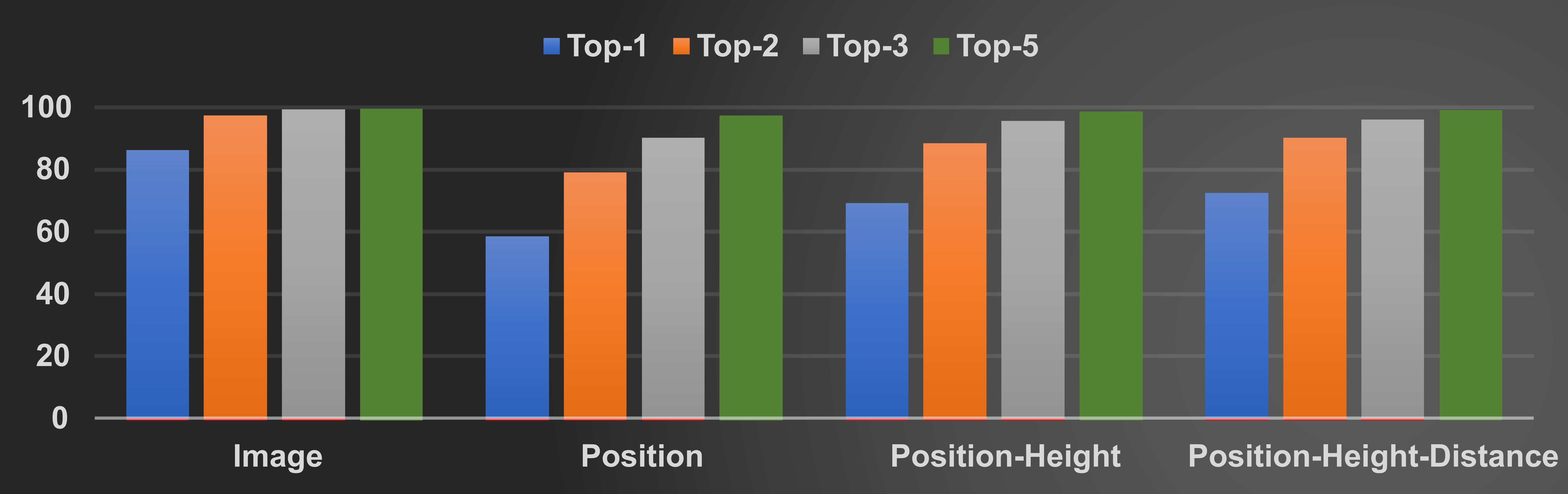}
	\caption{This figure plots the top-k accuracies $(k \in (1,2,3,5))$ for the proposed sensing-aided beam prediction solution. It is observed the vision-aided beam prediction solution outperforms the other approaches. }
	\label{fig:beam_pred_acc}
\end{figure}

\subsection{Numerical Results:} \label{sec:results}
With the experimental setup described in Section~\ref{sec:exp_setup}, in this subsection, we study the beam prediction performance of the proposed solution. In the first set of studies, we evaluate the performance of the proposed solutions from a machine learning perspective, i.e., beam prediction accuracy per approach, number of samples required for training, etc.

\textbf{Beam Prediction Accuracy Comparison:} The UE in this study is a drone, which brings its own set of challenges, such as the six degrees of freedom in motion, the variability of orientation, etc. In Fig.~\ref{fig:beam_pred_acc} we compare the performance of the different proposed solutions for the mmWave drone beam prediction task. It is observed in Fig.~\ref{fig:beam_pred_acc} that the position-alone approach achieves only $\approx 59\%$ top-1 accuracy. This is an interesting result as \textbf{ it highlights that for mmWave communication using drones, position alone might not be sufficient in predicting the optimal beam indices.} The combined modalities achieve an improvement of $\approx 10 - 14\%$ over the position-alone beam prediction solution. These results highlight the need for additional sensory data such as the height and distance of the drone from the basestation. Images can successfully capture the orientation and location of the object in the visual field. This is reflected in the performance of the vision-aided solution; it achieves a top-1, top-3, and top-5 accuracy of $86.32\%$, $99.41\%$, and $99.69\%$.


\begin{figure}[t]
	\centering
	\subfigure[]{\centering \includegraphics[width=1\columnwidth]{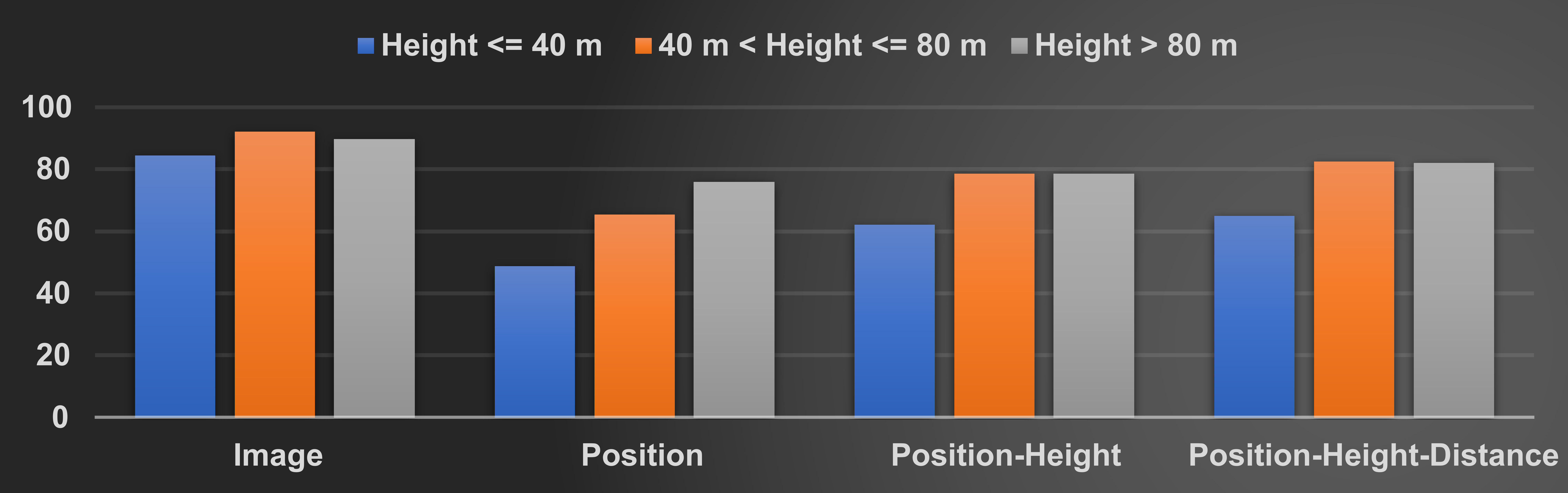}\label{fig:height_vs_acc_comp}}
	\subfigure[]{\centering \includegraphics[width=1\columnwidth]{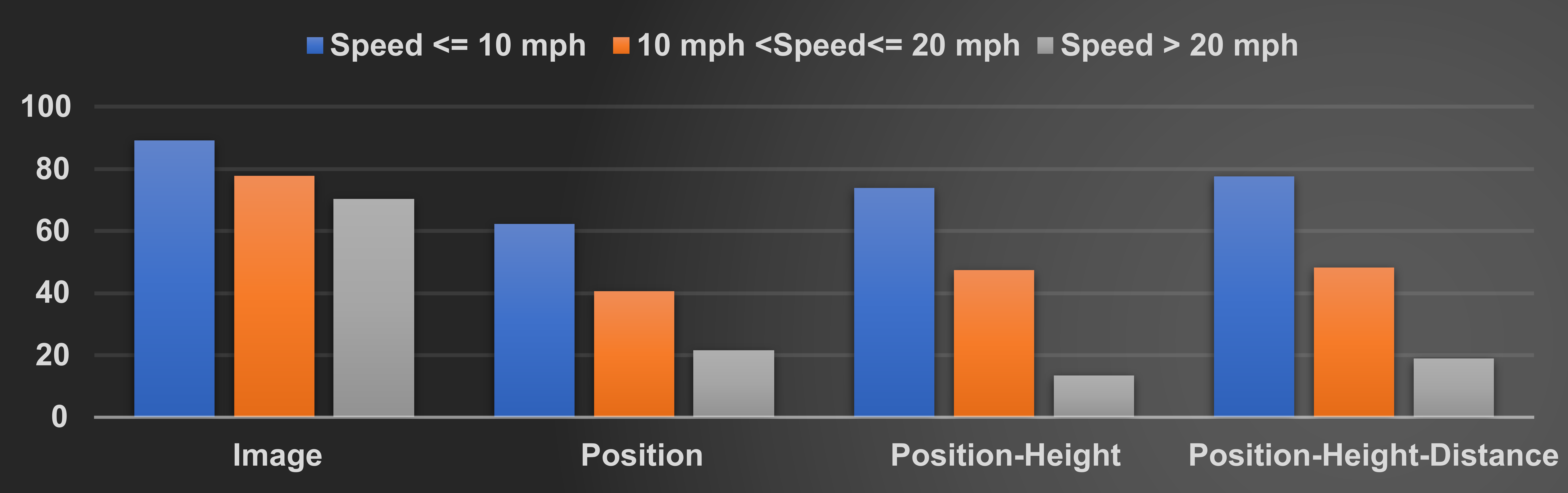}\label{fig:speed_vs_acc_com}}
	\caption{This figures studies the impact of speed and height on the beam prediction performance of the proposed solutions. }
	\label{fig:metric_comp}
	\vspace{-4mm}
\end{figure}


\textbf{Impact of height and speed on beam prediction accuracy:} As observed previously, the beam prediction accuracy improved significantly from the position-only solution with additional sensing data such as the height and the distance of the drone. Here, we consider two important data modalities, i.e., (i) height and (ii) speed, of the drone and study the impact of these data modalities on beam prediction accuracy. For both the data modalities, we first divide the test dataset into three sub-groups. For example, for the accuracy analysis based on speed, we divide the dataset into three groups: slow, medium, and fast.  Next, we calculate the beam prediction accuracy for each of the sub-groups. In Fig.~\ref{fig:height_vs_acc_comp} and Fig.~\ref{fig:speed_vs_acc_com}, we present the beam prediction accuracy versus the height and speed of the drone, respectively. Fig.~\ref{fig:height_vs_acc_comp} highlights an interesting fact that all the four ML-based solution makes the most mistakes in prediction when the drone is flying low, i.e., the height is less than 40 meters. For the speed-based analysis, we observe that the beam prediction performance starts degrading for higher traveling speeds of the drone.

\section{Conclusion}\label{sec:conc}
This paper develops a novel approach that leverages sensory data, such as visual and position data, for fast and accurate beam prediction in mmWave/THz drone communication systems. To evaluate the efficacy of the proposed solution, we adopt a real-world multi-modal drone communication scenario from the DeepSense 6G dataset. We perform an in-depth evaluation of different sensory modalities and compare the impact of different sensing data on the beam prediction accuracy. The proposed vision-aided solution achieves top-1 and top-5 accuracies of $86.32\%$ and $99.69\%$, respectively. This highlights the promising gains of leveraging sensory data to reduce the beam training overhead in mmWave/THz drone communication systems.


\end{document}